\documentclass{aipproc}

\begin{document}

\title{functionalObjects.h: Using Symbolic Syntax in C++ Programs}
\author{R. Nolty}
\affiliation{California Institute of Technology, Pasadena, California
  91125}

\begin{abstract}
  functionalObjects.h allows the C++ programmer performing common
  mathematical calculations to use a more symbolic syntax rather than
  an algorithmic syntax.  This is not as ambitious as a symbolic
  manipulation program such as Mathematica; it is more like having the
  ability to drop a very simple Mathematica statement into a C++
  program.
\end{abstract}

\maketitle

\section{Introduction}

A physicist is often faced with the task of writing a program to
perform a relatively straightforward mathematical manipulation.  For
example, she may need to multiply a couple of multivariate functions
together and integrate over one variable to obtain a new multivariate
function.  Using FORTRAN or C or a procedural approach with C++, the
resulting code may be several hundred lines long and include calls to
CERNlib routines with non-obvious names and calling sequences.

Similarly, a physicist may be reading a piece of procedural code
written by someone else, and only after several hours of study be able
to verify that the code is indeed performing a simple mathematical
function.

Computer Algebra Systems (CAS) such as Mathematica or Maple avoid
these problems; they allow the physicist to express the mathematical
function to be evaluated rather than the algorithm for evaluating it.
However, most of us have most of our analysis paraphernalia and
infrastructure in programming languages.  Results that are easily
obtained in a CAS are often not useful in solving an analysis
problem.

This paper documents an initial attempt to use the facilities of C++
to allow a programmer to express mathematical operations more directly
in the programming language itself.

\section{What does it look like?}

A program to calculate a double definite integral, 

\[ \int_0^1 dy\  \int_0^y dx\  x*y \]

\noindent
could be written like this:

\begin{verbatim}
#include <stl.h>
#include "functionalObjects.h"

functionalObjectGlobals theGlobals;

main()
{
  char* x = "x";
  char* y = "y";
  theGlobals.registerArgument(x);
  theGlobals.registerArgument(y);

  cout << "Integral: " <<
    Evaluate(Integrate(y, 0.0, 1.0,
      Multiply(y, 
        Integrate(x, 0.0, y, x)))) << 
          endl;
}
\end{verbatim}

All the action is in the last statement.  Reading it from the inside
out, Integrate(x,0,y,x) is a C++ function returning a functionalObject
which represents the definite integral over dx, with lower limit 0 and
upper limit y, of the function x.  Multiply(y, Integrate(...)) is a C++
function returning a functionalObject which represents the product of
the function y, and the function defined by Integrate(...).
Integrate(y,0,1,Multiply(...)) is a C++ function returning a
functionalObject which represents the definite integral over dy, with
lower limit 0 and upper limit 1, of the function represented by
Multiply(...).  Finally, Evaluate(...) is a C++ function which asks
its argument, a functionalObject, to evaluate its numeric value and
returns that value.  Thus, if the program is run, it produces the
output

\begin{verbatim}
Integral: 0.124994
\end{verbatim}

\noindent
which differs by roundoff error from the exact result 1/8.

\section{Distinctives of this effort}

The effort to date is very preliminary; it has involved only a few
days of thinking, a couple of days of coding, and a few hundred lines
of C++.

Although functionalObjects.h allows symbolic functions to be
expressed, it is not a symbolic manipulation program.  For example,
the product of one functionalObject representing the function $x$, and
another $1/x$, would not be simplified to 1.  Instead, $x$ would be
evaluated, $1/x$ would be evaluated, and their product would be
evaluated.  The system is designed to produce numeric results, not
symbolic results.

The chief aim is to produce programs that are easy to write, easy to
read, and easy to maintain.

\section{Implementation}

The heart of the package is the abstract base class functionalObject,
which simply defines a virtual function evaluate(), returning a
double.  While the evaluate() function takes no arguments, the
numerical value of a function may depend on a functionalArgument.  For
example, a functionalObject may represent sin(x).  If so, the
evaluate() function will check the current value of x, and return its
sine.  functionalArguments are managed by a global structure,
functionalObjectGlobals.  It has functions to declare an argument, to
set the value of an argument, and to inquire the current value of an
argument.

Often, more than one function may depend on the same argument.  For
example, a neutrino cross section and a neutrino flux may depend on
the same argument, Enu.  If they are evaluated at the same time, they
will both query the functionalObjectGlobals for the current value of
Enu (and any other arguments they depend on).

Application programmers may define classes inheriting from
functionalObject to compute arbitrarily complex functions.  The
package also provides a few commonly-needed functions; for example, a
multiplyObject is a class implementing the product of two functions
(each represented by a functionalObject).  It simply stores pointers
to the two functionalObjects.  When the multiplyObject is evaluated,
it evaluates its two functions (actually it asks the two
functionalObjects to evaluate themselves) and returns the product.

An integrateObject represents a definite integral.  When it is asked
to evaluate itself, it varies its integration variable (by
communication with the functionalObjectGlobals) over its integration
range and at each point evaluates its argument, which must be a
functionalObject, until it has computed the definite integral.

Programs are made a bit more readable by the existence of certain
functions that construct and return functionalObjects.  For example,
Multiply(functionalObject fcn1,functionalObject fcn2) constructs and
returns a multiplyObject.  Some overloads also implicitly create very
simple functions.  Mutliply(3.0,fcn1) creates a doubleObject which
always evaluates to the double 3.0, and then returns a multiplyObject
that represents the product of the new doubleObject and fcn1.
Multiply(``x'',fcn1) creates an argumentObject that always evaluates
to the current value of the functionalArgument x, and then returns a
multiplyObject that represents the product of the new argumentObject
and fcn1.  Similarly, Integrate(char* integrationVariable,
functionalObject* lowerLimit, functionalObject* upperLimit,
functionalObject* integrand) constructs and returns the appropriate
integrateObject representing a definite integral.

\section{Shortcomings}

As stated above, the effort is not very mature at this point.  It
exhibits several shortcomings, some of which could be overcome with
further effort.

For some problems, a procedural approach could take advantage of
peculiarities of the problem to make a much more efficient algorithm.
The emphasis here is on ease of programming and ease of reading and
maintaining programs, not on efficiency.

In the current implementation, functions that implicitly depend on
arguments access their arguments by name.  It is a bit like the early
days of programming before formal parameters were invented.  A
function, squareX, that computes the square of x, is of no value if
you want to compute the square of y.  This greatly limits the ability
of a programmer to develop generic functions or use functions
developed by other programmers.  This limitation could be overcome by
making the evaluate() function accept arguments, which would be the
names of functionalArguments on which the function is to depend.

No thought has been given to memory management.  Some functions
implicitly declare new variables on the heap, but they are never
recovered when the functions go out of scope.

The operation of the package could be made more transparent.  For
example, an argument class could be declared whose constructor would
declare the argument to the functionalObjectGlobals.  Rather than
using the syntax Multiply(fcn1,fcn2), an overload of the operator*
could be used.  However, I am somewhat reluctant to make these
changes.  They would make the program appear simpler but it would not
really be any simpler.  I find that physicists are more comfortable
when they can see how the package works, rather than having the
mechanism hidden by programming gimmicks.

\section{Conclusion}

This effort, while very preliminary, shows that C++ has powerful
facilities making it possible to express some mathematical operations
more directly than has been possible in procedural languages.

The library and some example programs are available at

\ 

\noindent
http://www.hep.caltech.edu/\~nolty/functionalObjects/
functionalObjects.tgz

\ 

\noindent
in a gzipped tar archive.  No license has been developed, but if there
is interest I will place it under the Gnu Public License (GPL).

\nocite{*}
\bibliographystyle{aipproc}
\bibliography{test}

\end{document}